# Grover search algorithm

*Eva Borbely*

Technological University Budapest, Hungary
PhD student

**Introduction**

A quantum algorithm is a set of instructions for a quantum computer, however, unlike algorithms in classical computer science their results cannot be guaranteed. A quantum system can undergo two types of operation, measurement and quantum state transformation, operations themselves must be unitary (reversible). Most quantum algorithms involve a series of quantum state transformations followed by a measurement. Currently very few quantum algorithms are known and no general design methodology exists for their construction.

Almost all quantum algorithms can be expressed in a in a query (oracle) model where the input is given by a black box which answers queries of a certain form.

In the query model, the input $x_1,...,x_n$ is contained in a black box and can be accessed by queries to the black box. In each query we give $i$ to the black box, and the black box outputs $x_i$. The goal is to solve the problem with the minimum number of queries.

There are two ways to define the query box in the quantum model.

In the first case it has two inputs $i$ and $j$, where $i$ consisting of $\log_2 N$ bits and $j$ consisting of 1 bit. If the input to the query box is a basis state $|i\rangle|j\rangle$, the output is $|i\rangle|j \oplus x_i\rangle$ where $\oplus$ denotes addition modulo 2. If the input is a superposition : $\sum_{i,j} a_{i,j}|i\rangle|j\rangle$ , the output is $\sum_{i,j} a_{i,j}|i\rangle|j \oplus x_i\rangle$.

In the second form of quantum query the black box has just one input $i$. If the input is a state $\sum_i a_i |i\rangle$, the output is $\sum_i a_i (-1)^{x_i} |i\rangle$.

(This query model for example serves as a basis of Grover's search algorithm. )

A large class of problems can be specified as search problems of the form: "find some $x$ such that statement $f(x)$ is true."

Such problems range from database search to sorting. A sorting problem can be viewed as a search for a permutation for which the statement "the permutation $x$ takes the initial state to the desired sorted state" is true.

An *unstructured* search problem is one where nothing is known (or no assumption are used) about the structure of the solution space and the statement *f*. For example, determining $f(x_0)$ provides no information about the possible value of $f(x_1)$ for $x_0 \neq x_1$.

A *structured* search problem is one where information about the search and statement *f* can be exploited. For instance, searching an alphabetical list is a structured search problem and the structure can be exploited to construct efficient algorithms.

Consider the problem of searching for a phone number in an unsorted directory with N names.. This is an example of unstructured search problem. In order to find someone's phone number with a probability of ½, any classical algorithm will need to look at minimum of N/2 names. In the general case of an unstructured problem, randomly testing the truth of some statement $f(x_i)$ one by one is the best that can be done classically.

For a search space of size N, the general unstructured search problem requires O(N) evaluation of *f*.

On a quantum computer, however, Grover showed that the unstructured search problem can be solved with bounded probability within $O(\sqrt{N})$ evaluation of *f*.

The problem is to search trough a search space of N elements (for convenience we assume $N = 2^n$) and the search problem has exactly M solutions with $1 \leq M \leq N$.

In our case the search problem can be represented by a function *f*, wich takes as input register x, x =0,1,…, N-1. By definition $f(x) = 1$ if x is a solution to the search problem, and $f(x) = 0$ if x is not a solution to the search problem.

Classically we need N queries to solve this problem. If we are using probabilistic classical computer, we can reduce it to $\frac{N}{2}$ expected steps.

Grover showed that there is a quantum algorithm that solves this problem with $O(\sqrt{N})$ queries.

The simplest case is if M=1, namely there exist exactly one x such that $f(x) = 1$. In this case we consider the following problem:

- Label the records of the database with the integers 0, 1, 2, . . . , N -1 ,
- denote the unknown marked record by ω
- let *f* an *n* bit binary function $f : \{0,1\}^n \rightarrow \{0,1\}$

$$f(x) = \begin{cases} 1, & x = \omega \\ 0, & otherwise \end{cases}$$

*Search Problem for an Unstructured Database.* Find the record ω with the minimum amount of computational work, i.e., with the minimum number of queries.

To solve the Grover's search problem suppose we are supplied with a quantum oracle with the ability to recognize solution to the search problem.

The oracle is a unitary operator U, defined by its action on the computational basis:

$$U : |x\rangle|q\rangle \to |x\rangle|q \oplus f(x)\rangle$$

where $|x\rangle$ is the input register, $|q\rangle$ is the oracle qubit, and $\oplus$ denotes addition modulo 2. The oracle qubit $|q\rangle$ is a single qubit which is flipped if $f(x) = 1$, and is unchanged otherwise. We can check whether $x$ is a solution to our search problem by preparing $|x\rangle|0\rangle$, applying the oracle and checking to see if the oracle qubit has been flipped to $|1\rangle$.

It is useful to choose the state of the single – qubit register to be: $\frac{1}{\sqrt{2}}(|0\rangle - |1\rangle)$. We achieve this state simply applying the Hadamard transform to the $|1\rangle$ quantum state:

$$H \equiv \frac{1}{\sqrt{2}}\begin{bmatrix} 1 & 1 \\ 1 & -1 \end{bmatrix}, \quad \begin{matrix} |0\rangle \\ |1\rangle \end{matrix} \xrightarrow{H} \begin{cases} \frac{1}{\sqrt{2}}(|0\rangle + |1\rangle) \\ \frac{1}{\sqrt{2}}(|0\rangle - |1\rangle) \end{cases}$$

Now we apply the oracle to the new state:

$$U|x\rangle\left(\frac{|0\rangle - |1\rangle}{\sqrt{2}}\right) \to (-1)^{f(x)}|x\rangle\left(\frac{|0\rangle - |1\rangle}{\sqrt{2}}\right).$$

For instance, if *x* is a solution to the search problem then the final state will be:

$$-|x\rangle\left(\frac{|0\rangle - |1\rangle}{\sqrt{2}}\right).$$

It is obvious, that the state of the oracle qubit is not changed, so we may ignore it, and obtain:

$$|x\rangle \xrightarrow{U} (-1)^{f(x)}|x\rangle.$$

It is easy to see, that the oracle transformation U is equivalent with the following transformation: $U = I - 2|\omega\rangle\langle\omega|$, where $|\omega\rangle$ is the only solution of the search problem, and I is the identity matrix.

If $|x\rangle = |\omega\rangle$, then: $U|\omega\rangle = I|\omega\rangle - 2|\omega\rangle\langle\omega||\omega\rangle = I|\omega\rangle - 2|\omega\rangle = -|\omega\rangle$.

If $|x\rangle \neq |\omega\rangle$ then we have: $U|x\rangle = I|x\rangle - 2|\omega\rangle\langle\omega\|x\rangle = I|x\rangle = I|x\rangle$ since ω and $x$ are orthogonal to each other if $\omega \neq x$.

Thus the oracle U flips the sign of $|x\rangle$ if $|x\rangle = |\omega\rangle$ but operates trivially on all $|x\rangle \neq |\omega\rangle$. More precisely, when U operates on some vector $|x\rangle$ in Hilbert space $H^{2^n}$ ($N = 2^n$), reflects it about the hyperplane orthogonal to $|\omega\rangle$. We know that the reflection is performed for some basis state $|\omega\rangle$, but nothing is known about the string ω itself. Our task is to determine ω. To achieve our aim, we are using the so – called *Grover iteration*:

1. Create a perfect random state $|\Psi\rangle = \frac{1}{\sqrt{2^n}} \sum_{i=0}^{2^n-1} |x_i\rangle$ by application of the Hadamard transformation on the state $|0\rangle^{\otimes n}$.

We know about $|\omega\rangle$ that it is a base state, which means that $\langle\Psi\|\omega\rangle = \frac{1}{\sqrt{2^n}}$. Measuring the state $|\Psi\rangle$ by projection onto its base $\{|x\rangle\}$ would return the state $|\omega\rangle$ only with probability $\frac{1}{N}$.

2. Combine the unknown reflection U with some known reflection $V = 2|\Psi\rangle\langle\Psi| - I$. It flips the sign of $|x\rangle$ if it is orthogonal to $|\Psi\rangle$ and preserves the sign of $|x\rangle$, if $|x\rangle = |\Psi\rangle$:

    — $V|x\rangle = 2|\Psi\rangle\langle\Psi|x\rangle| - I|x\rangle = -|x\rangle$ because $\langle\Psi\|x\rangle = 0$
    — $V|\Psi\rangle = 2|\Psi\rangle\langle\Psi\|\Psi\rangle - I|\Psi\rangle = |\Psi\rangle$, since $\langle\Psi\|\Psi\rangle = I$.

Acting on some arbitrary vector, V preserves the component along $|\Psi\rangle$ but flips all components in the hyperplane orthogonal to $|\Psi\rangle$.

3. Apply the operator **G=V*U** to some vector $|x\rangle$ with the following effect:

    — the vector $|\Psi\rangle$ and $|\omega\rangle$ span a plane in $H^{\otimes N}$ with a vector $|\omega^\perp\rangle$ that is orthogonal to $|\omega\rangle$ in that plane;
    — any vectors $|x\rangle$ in that plane is by U flipped about $|\omega^\perp\rangle$ and by V flipped about $|\Psi\rangle$.

To understand what this means, consider the geometry in a two-dimensional Hilbert space:

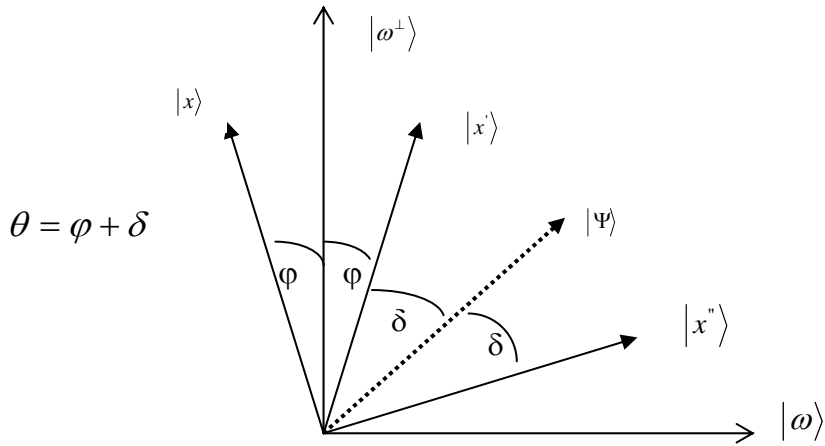

— the angle between $|\Psi\rangle$ and $|\omega\rangle$ is given as:

$$\langle \Psi \| \omega \rangle = \frac{1}{\sqrt{N}} = \frac{1}{\sqrt{2^n}} = \cos\left(\frac{\pi}{2} - \theta\right) = \sin\theta$$

— a vector $|x\rangle$ is by U flipped about $|\omega^{\perp}\rangle$ and thereby changed its angle by $2\varphi$

— the vector $U|x\rangle = |x'\rangle$ is by V flipped about $|\Psi\rangle$, thus changing its angle by $2\delta$

— the total change of angle effected by **G=V*U** is thus $2\varphi + 2\delta = 2\theta$, which is the effect of one Grover iteration.

*Remark:*

He Grover iteration may be seen to be a consequence of the following elementary theorem of 2-dimensional real Euclidian geometry.

<u>Theorem:</u> Let *L* and *M* be two mirror lines in the Euclidian plan **R**² intersecting at a point *O* and let $\alpha$ be the angle in the plane from *L* to *M*. Then the operation of reflection in *L* followed by reflection in *M* is just a rotation by angle $2\alpha$ about the point *O*.

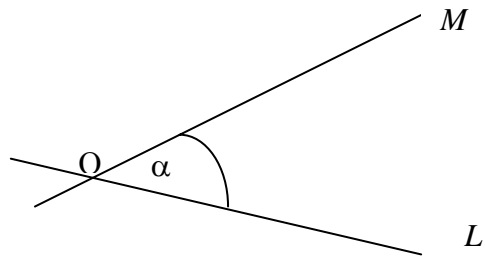

4. Repeat the Grover iteration until the possibility the input item in the data base approaches the value 1.

We assume that we need some $k$ iteration to bring $|\Psi\rangle$ close to $|\omega\rangle$ ( or to turn it away from $|\omega^\perp\rangle$ by an angle of $\frac{\pi}{2}$.

After $k$ iteration we thus get a rotation by $\theta + 2k\theta = \frac{\pi}{2}$.

Since $\sin\theta \cong \theta = \frac{1}{\sqrt{N}}$ for large $N$, we iterate until:

$$(2k+1)\theta \cong \frac{\pi}{2}, \Rightarrow k = \text{round}\left(\frac{\pi}{4\theta} - \frac{1}{2}\right)$$

$$\text{or } k \cong \frac{\pi}{4}\sqrt{N}.$$

Only $\frac{\pi}{4}\sqrt{N}$ iteration are required to find $|\omega\rangle$ with high probability.

For example assuming that the data base contains just 4 items. A classical sequential search would requires 2 steps to find a particular item. A quantum-mechanical search for an item $|x\rangle$ would set out $|\langle x|\langle\omega|\rangle| = \frac{1}{\sqrt{4}} = \frac{1}{2} = \sin\theta$ which means that $\theta = \sin^{-1}\frac{1}{2} = 30°$. A rotation by $2\theta = 60°$ brings $|x\rangle$ into a $90°$ angle with $|\omega^\perp\rangle$, or in line with $|\omega\rangle$.

**Multiple solutions**

Suppose that the search problem has exactly M solutions, with $1 \leq M \leq N$, $N = 2^n$ and $M$ is known. In this case the oracle introduces a reflection in the hyperplane orthogonal to the vector $|\beta\rangle = \frac{1}{\sqrt{M}}\sum_{i=1}^{M}|\omega_i\rangle$, or in other manner $|\beta\rangle = \frac{1}{\sqrt{M}}\sum_{x=f^{-1}(1)}|x\rangle$, the equal weighted superposition of the marked computational basis states.

The original state $|\Psi\rangle = \frac{1}{\sqrt{N}}\sum_{i=1}^{N-1}|x\rangle$ can be rewritten as:

$$|\Psi\rangle = \frac{\sqrt{N-M}}{\sqrt{N}}\left(\frac{1}{\sqrt{N-M}}\sum_{x=f^{-1}(0)}|x\rangle\right) + \frac{\sqrt{M}}{\sqrt{N}}\left(\frac{1}{\sqrt{M}}\sum_{x=f^{-1}(1)}|x\rangle\right),$$

or

$$|\Psi\rangle = \sqrt{\frac{N-M}{N}}|\alpha\rangle + \sqrt{\frac{M}{N}}|\beta\rangle,$$

where $|\alpha\rangle \equiv \frac{1}{\sqrt{N-M}}\sum_{x=f^{-1}(0)}|x\rangle$, $|\beta\rangle \equiv \frac{1}{\sqrt{M}}\sum_{x=f^{-1}(1)}|x\rangle$.

Using the simplest notation, $|\Psi\rangle = \cos\theta|\alpha\rangle + \sin\theta|\beta\rangle$.

Geometrical representation:

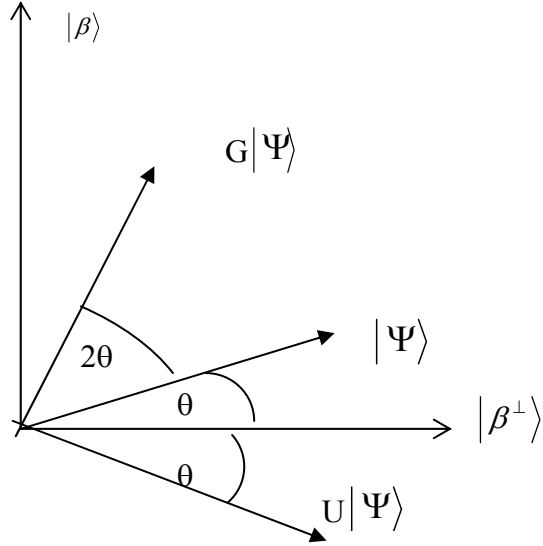

The effect of **G = V*U** is the following:

— the oracle U performs a reflection about the vector $|\alpha\rangle$ which is orthogonal to $|\beta\rangle$:
$$U(\cos\theta|\alpha\rangle + \sin\theta|\beta\rangle) = \cos\theta|\alpha\rangle - \sin\theta|\beta\rangle$$

— the vector $\cos\theta|\alpha\rangle - \sin\theta|\beta\rangle$ is by $V = 2|\Psi\rangle\langle\Psi| - I$ flipped about $|\Psi\rangle$

— the product of two reflection is a rotation and after $k$ iteration the state:
$$G^k|\Psi\rangle = \cos(2k+1)\theta|\alpha\rangle + \sin(2k+1)\theta|\beta\rangle$$

For large N, when $M \ll N$ we have $\theta \approx \sin\theta \approx \sqrt{\dfrac{M}{N}}$ and:

$$(2k+1)\theta = \frac{\pi}{2} ,$$

$$(2k+1)\sqrt{\frac{M}{N}} = \frac{\pi}{2} \Rightarrow k = \text{round}\left(\frac{\pi}{4}\sqrt{\frac{N}{M}} - \frac{1}{2}\right).$$

So the state is close to $|\beta\rangle$ after a number of iteration $k \approx \dfrac{\pi}{4}\sqrt{\dfrac{N}{M}}$ with probability:

$$prob = |\sin^2(2k+1)\theta|.$$

Reverting to the previous example; the probability to find one marked item from 4 is $prob = |\sin^2(2k+1)\theta| = |\sin^2 3\cdot\theta| = |\sin^2 3\cdot 30°| = 1$.

**Quantum search algorithm in detail:** $(M = 1, N = 2^n)$

*Input:*

— $x_1 = \{0,1\}, x_2 = \{0,1\}, \cdots, x_N = \{0,1\}$, $f : \{0,1\}^N \to \{0,1\}$; $f(x) = 0$ for all $0 \leq x \leq N$ except $\omega = x_i$, for which $f(x_i) = 1$

— $n$ qubits in the state $|0\rangle$ and 1 qubit, the oracle qubit in state: $\left[\dfrac{|0\rangle - |1\rangle}{\sqrt{2}}\right]$.

*Procedure:*

1. Initial sate: $|0\rangle^{\otimes n}\left[\dfrac{|0\rangle - |1\rangle}{\sqrt{2}}\right]$

2. Apply the $H^{\otimes n}$ to the first $n$ qubits: $\Rightarrow \dfrac{1}{\sqrt{N}}\sum_{x=0}^{N-1}|x\rangle\left[\dfrac{|0\rangle - |1\rangle}{\sqrt{2}}\right]$

3. Apply the Grover iteration $k \approx \dfrac{\pi}{4}\sqrt{N}$ times

$\Rightarrow [(2|\Psi\rangle\langle\Psi| - I)(I - 2|\omega\rangle\langle\omega|)]^k \dfrac{1}{\sqrt{N}}\sum_{x=0}^{N-1}|x\rangle\left[\dfrac{|0\rangle - |1\rangle}{\sqrt{2}}\right] \approx |\omega\rangle\left[\dfrac{|0\rangle - |1\rangle}{\sqrt{2}}\right]$

4. Measure the first $n$ qubits $\Rightarrow |\omega\rangle$.

**Inversion about the average**

There is an alternative way to visualize the Grover iteration that is sometimes useful, as an "inversion about the average".

The unitary transformation $D_n : \sum_{i=0}^{N} a_i |x_i\rangle \to \sum_{i=0}^{N} (2A - a_i)|x_i\rangle$, where $A$ is the average of $\{a_i | 0 < i < N\}$ can be performed by the matrix:

$$D = \begin{bmatrix} \frac{2}{N}-1 & \frac{2}{N} & \cdots & \frac{2}{N} \\ \frac{2}{N} & \frac{2}{N}-1 & \cdots & \frac{2}{N} \\ \vdots & \vdots & \vdots & \vdots \\ \frac{2}{N} & \frac{2}{N} & \cdots & \frac{2}{N}-1 \end{bmatrix},$$

or simply: $D_{i,j} = \begin{cases} \frac{2}{N}, & \text{if } i \neq j \\ \frac{2}{N}-1, & \text{if } i = j \end{cases}$

$D$ has two properties:
1. it is unitary
2. it can be seen as "inversion about average".

Proof:
1. For $N = 2^n$, operator $D$ can be decomposed and rewritten as:

$$D = H^{\otimes n} \begin{bmatrix} 1 & 0 & \cdots & 0 \\ 0 & -1 & \cdots & 0 \\ \vdots & \vdots & \vdots & \vdots \\ 0 & 0 & \cdots & -1 \end{bmatrix} H^{\otimes n} = H^{\otimes n} \left[ \begin{bmatrix} 2 & 0 & \cdots & 0 \\ 0 & 0 & \cdots & 0 \\ \vdots & \vdots & \vdots & \vdots \\ 0 & 0 & \cdots & 0 \end{bmatrix} - I \right] H^{\otimes n} =$$

$$= H^{\otimes n} \begin{bmatrix} 2 & 0 & \cdots & 0 \\ 0 & 0 & \cdots & 0 \\ \vdots & \vdots & \vdots & \vdots \\ 0 & 0 & \cdots & 0 \end{bmatrix} H^{\otimes n} - I = H^{\otimes n} \begin{bmatrix} \frac{2}{\sqrt{N}} & \frac{2}{\sqrt{N}} & \cdots & \frac{2}{\sqrt{N}} \\ 0 & 0 & \cdots & 0 \\ \vdots & \vdots & \vdots & \vdots \\ 0 & 0 & \cdots & 0 \end{bmatrix} - I =$$

$$= \begin{bmatrix} \frac{2}{\sqrt{N}} & \frac{2}{\sqrt{N}} & \cdots & \frac{2}{\sqrt{N}} \\ \frac{2}{\sqrt{N}} & \frac{2}{\sqrt{N}} & \cdots & \frac{2}{\sqrt{N}} \\ \vdots & \vdots & \vdots & \vdots \\ \frac{2}{\sqrt{N}} & \frac{2}{\sqrt{N}} & \cdots & \frac{2}{\sqrt{N}} \end{bmatrix} - I = \begin{bmatrix} \frac{2}{\sqrt{N}}-1 & \frac{2}{\sqrt{N}} & \cdots & \frac{2}{\sqrt{N}} \\ \frac{2}{\sqrt{N}} & \frac{2}{\sqrt{N}}-1 & \cdots & \frac{2}{\sqrt{N}} \\ \vdots & \vdots & \vdots & \vdots \\ \frac{2}{\sqrt{N}} & \frac{2}{\sqrt{N}} & \cdots & \frac{2}{\sqrt{N}}-1 \end{bmatrix}$$

Observe that $D$ can be expressed as the product of three unitary matrices: two Hadamard matrices separated by a conditional phase shift matrix. Therefore D is also unitary.

2. Let $|\Psi\rangle = \sum_{i=0}^{N-1} a_i |i\rangle$ be the state before $D$. Then, the state after $D$ is: $|\Psi'\rangle = \sum_{i=0}^{N-1} a_i' |i\rangle$, where $a_i' = \left(\frac{2}{N}-1\right)a_i + \sum_{i \neq j} \frac{2}{N} a_j$. We can rewrite this as $a_i' = -a_i + \sum_{j=0}^{N-1} \frac{2}{N} a_j$ and $a_i' + a_i = \sum_{j=0}^{N-1} \frac{2}{N} a_j$. Let $A = \sum_{j=0}^{N-1} \frac{1}{N} a_j$ be the average of amplitudes $a_i$. Then, we have $a_i' + a_i = 2A$ and, if $a_i = A + \Delta$, then $a_i' = A - \Delta$. Thus, the effect of the $D$ operator is that every amplitude $a_i$ is replaced by its reflection against the average of all $a_i$.

Geometrical representation:

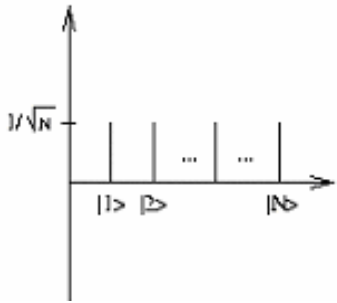

Amplitudes in initial state
(fig.1.)

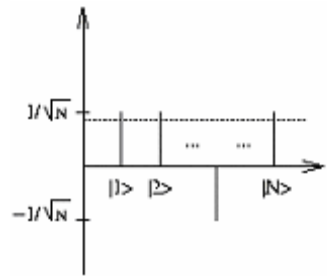

Amplitudes after the first oracle trnsformation: U
(fig.2.)

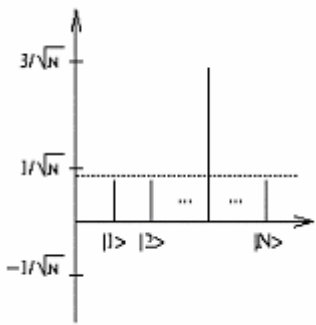

Amplitudes after $D$
(fig.3.)

As we can see in fig. 2., after the first oracle transformation $U = 2|\omega\rangle\langle\omega| - I$, the amplitude of $\omega = x_i = 1$ is $-\frac{1}{\sqrt{N}}$, and all the other amplitudes are $\frac{1}{\sqrt{N}}$. The average is:

$$\frac{(N-1)\cdot\frac{1}{\sqrt{N}} - \frac{1}{\sqrt{N}}}{N} = \frac{\frac{N}{\sqrt{N}} - \frac{2}{\sqrt{N}}}{N} = \frac{1}{\sqrt{N}} - \frac{2}{N\sqrt{N}} \approx \frac{1}{\sqrt{N}}$$ 

for large $N$. Therefore, after applying $D$, the amplitude of $|i\rangle$ with $x_i = 1$ becomes almost $\frac{3}{\sqrt{N}}$ and the amplitudes of all other basis states $|j\rangle$ slightly less than $\frac{1}{\sqrt{N}}$ (fig.3.).

The next query (oracle transformation) makes the amplitude of $|i\rangle$ with $x_i = 1$ approximately $-\frac{3}{\sqrt{N}}$. The average of all amplitudes is: 

$$\frac{(N-1)\cdot\frac{1}{\sqrt{N}} - \frac{3}{\sqrt{N}}}{N} = \frac{1}{\sqrt{N}} - \frac{4}{N\sqrt{N}} \approx \frac{1}{\sqrt{N}}$$

, which is slightly less than $\frac{1}{\sqrt{N}}$, and reflecting against it make the amplitude $|i\rangle$ with $x_i = 1$ almost $\frac{5}{\sqrt{N}}$.

A precise calculation made by Grover shows that after $\frac{\pi}{4}\sqrt{N}$ steps the amplitude of $|i\rangle$ with $x_i = 1$ is $1 - o(1)$. Therefore, the measurement gives the correct answer with probability $1 - o(1)$.

**Example for N = 4 (matrix representation)**

In this case we need $n=2$ qubits and another one with 1 qubit, the oracle qubit. We need just at $k = \text{round}\left(\frac{\pi}{4}\sqrt{N}\right) = \text{round}\left(\frac{\pi}{2}\right) = 1$ oracle query to resolve the search problem.

The initial state is $|\Psi_0\rangle = |00\rangle$, and after applying the Hadamard gates we get:

$$|\Psi\rangle = H^{\otimes 2}|00\rangle = (H|0\rangle)^{\otimes 2} = \left[\frac{1}{\sqrt{2}}\begin{pmatrix}1 & 1 \\ 1 & -1\end{pmatrix}\begin{pmatrix}1 \\ 0\end{pmatrix}\right]^{\otimes 2} = \left[\frac{1}{\sqrt{2}}\begin{pmatrix}1 \\ 1\end{pmatrix}\right]^{\otimes 2} = \frac{1}{2}\begin{pmatrix}1 \\ 1\end{pmatrix} \otimes \begin{pmatrix}1 \\ 1\end{pmatrix} = \frac{1}{2}\begin{pmatrix}1 \\ 1 \\ 1 \\ 1\end{pmatrix}$$

$$|\Psi\rangle = \frac{1}{2}\begin{pmatrix} 1 \\ 1 \\ 1 \\ 1 \end{pmatrix}$$

Suppose that $f(3)=1$ and $f(i)=0$ for $i \neq 3$.

Than the oracle transformation is the identity matrix with the 3$^{th}$ element of the diagonal equal to $-1$.

$$U = \begin{pmatrix} 1 & 0 & 0 & 0 \\ 0 & 1 & 0 & 0 \\ 0 & 0 & 1 & 0 \\ 0 & 0 & 0 & -1 \end{pmatrix}$$

In the next step we apply the oracle transformation $U$ to the state $|\Psi\rangle = \frac{1}{2}\begin{pmatrix} 1 \\ 1 \\ 1 \\ 1 \end{pmatrix}$:

$$|\Psi_1\rangle = U|\Psi\rangle = \begin{pmatrix} 1 & 0 & 0 & 0 \\ 0 & 1 & 0 & 0 \\ 0 & 0 & 1 & 0 \\ 0 & 0 & 0 & -1 \end{pmatrix} \frac{1}{2}\begin{pmatrix} 1 \\ 1 \\ 1 \\ 1 \end{pmatrix} = \frac{1}{2}\begin{pmatrix} 1 \\ 1 \\ 1 \\ -1 \end{pmatrix}.$$

We have seen that the $V = 2|\Psi\rangle\langle\Psi| - I$ operator is equivalent with the so-called "diffusion operator" $D = \begin{cases} \dfrac{2}{N}-1 & \text{if } i=j \\ \dfrac{2}{N} & \text{if } i \neq j \end{cases}$ or in this case: $D = \begin{bmatrix} -\dfrac{1}{2} & \dfrac{1}{2} & \dfrac{1}{2} & \dfrac{1}{2} \\ \dfrac{1}{2} & -\dfrac{1}{2} & \dfrac{1}{2} & \dfrac{1}{2} \\ \dfrac{1}{2} & \dfrac{1}{2} & -\dfrac{1}{2} & \dfrac{1}{2} \\ \dfrac{1}{2} & \dfrac{1}{2} & \dfrac{1}{2} & -\dfrac{1}{2} \end{bmatrix}$

$$|\Psi 2\rangle = D|\Psi_1\rangle = \begin{bmatrix} -\dfrac{1}{2} & \dfrac{1}{2} & \dfrac{1}{2} & \dfrac{1}{2} \\ \dfrac{1}{2} & -\dfrac{1}{2} & \dfrac{1}{2} & \dfrac{1}{2} \\ \dfrac{1}{2} & \dfrac{1}{2} & -\dfrac{1}{2} & \dfrac{1}{2} \\ \dfrac{1}{2} & \dfrac{1}{2} & \dfrac{1}{2} & -\dfrac{1}{2} \end{bmatrix} \cdot \frac{1}{2}\begin{bmatrix} 1 \\ 1 \\ 1 \\ -1 \end{bmatrix} = \frac{1}{2}\begin{bmatrix} 0 \\ 0 \\ 0 \\ 2 \end{bmatrix} = \begin{bmatrix} 0 \\ 0 \\ 0 \\ 1 \end{bmatrix}.$$

According to the Grover iteration, the amplitude of the marked state increased by ½, while the amplitude of other states decreased to zero.

Now if we measure the final state $|\Psi_2\rangle$, then we obtain the correct answer with unit probability.

**Example for N = 8**

If $N = 2^3 \Rightarrow$ there are 3 qubits in the first register and 1 qubit in the second register, which is the oracle qubit. For N = 8, the operator G will be applied k = 2 times as follows from $k = \text{round}\left(\frac{\pi}{4}\sqrt{N}\right) = \text{round}(2,2214)$.

The initial state is $|\Psi_0\rangle = |000\rangle$.

After applying Hadamard gates, we get: $|\Psi\rangle = H^{\otimes 3}|000\rangle = (H|0\rangle)^{\otimes 3} = \frac{1}{\sqrt{8}}\sum_{i=0}^{7}|i\rangle$.

Suppose that we are searching for the element with index 5. Since $|5\rangle = |101\rangle$, after the $U$ oracle transformation we obtain: $U|101\rangle = -|101\rangle$, and $U|i\rangle = |i\rangle$, if $i \neq 5$.

Define $|s\rangle = \frac{1}{\sqrt{7}}\sum_{\substack{i=0 \\ i\neq 5}}^{7}|i\rangle = \frac{1}{\sqrt{7}}(|000\rangle + |001\rangle + |010\rangle + |011\rangle + |100\rangle + |110\rangle + |111\rangle)$.

Then $|\Psi\rangle = \frac{\sqrt{7}}{\sqrt{8}}|s\rangle + \frac{1}{\sqrt{8}}|101\rangle = \frac{\sqrt{7}}{2\sqrt{2}}|s\rangle + \frac{1}{2\sqrt{2}}|101\rangle$.

The value of $\theta$ is: $\theta = 2.\arccos\sqrt{\frac{7}{8}} \cong 41,4°$.

The next step is: $|\Psi_1\rangle = U|\Psi\rangle = \frac{1}{\sqrt{8}}(|000\rangle + |001\rangle + |010\rangle + |011\rangle + |100\rangle - |101\rangle + |110\rangle + |111\rangle)$

or using the state $|s\rangle$ we can write $|\Psi_1\rangle$ as: $|\Psi_1\rangle = |\Psi\rangle - \frac{1}{\sqrt{2}}|101\rangle$.

In the next step we have to apply the $2|\Psi\rangle\langle\Psi| - I$ operator:

$$|\Psi_2\rangle = (2|\Psi\rangle\langle\Psi| - I)\Psi_1\rangle = (2|\Psi\rangle\langle\Psi| - I)\left(|\Psi\rangle - \frac{1}{\sqrt{2}}|101\rangle\right) =$$

$$= 2|\Psi\rangle - \frac{2}{\sqrt{2}}|\Psi\rangle\langle\Psi||101\rangle - |\Psi\rangle + \frac{1}{\sqrt{2}}|101\rangle = 2|\Psi\rangle - \frac{2}{\sqrt{2}}\cdot\frac{1}{2\sqrt{2}}|\Psi\rangle - |\Psi\rangle + \frac{1}{\sqrt{2}}|101\rangle =$$

$$= \frac{1}{2}|\Psi\rangle + \frac{1}{\sqrt{2}}|101\rangle$$

or using $|\Psi\rangle = \frac{\sqrt{7}}{2\sqrt{2}}|s\rangle + \frac{1}{2\sqrt{2}}|101\rangle$,

$$|\Psi_2\rangle = \frac{1}{2}\left(\frac{\sqrt{7}}{2\sqrt{2}}|s\rangle + \frac{1}{2\sqrt{2}}|101\rangle\right) + \frac{1}{\sqrt{2}}|101\rangle = \frac{\sqrt{7}}{4\sqrt{2}}|s\rangle + \frac{5}{4\sqrt{2}}|101\rangle.$$

Let us confirm that the angle between $|\Psi\rangle$ and $|\Psi_2\rangle$ is $\theta$:

$$\cos\theta = \langle\Psi\|\Psi_2\rangle = \frac{1}{2}\langle\Psi\|\Psi\rangle + \frac{1}{\sqrt{2}}\langle\Psi\|101\rangle = \frac{1}{2} + \frac{1}{\sqrt{2}} \cdot \frac{1}{2\sqrt{2}} = \frac{3}{4}$$

$$\theta = \cos^{-1}\frac{3}{4} \cong 41,4°$$

which correspond with the previous result.

The second and last application of Grover iteration is similar:

$|\Psi_3\rangle$ is given by: $|\Psi_3\rangle = \frac{\sqrt{7}}{4\sqrt{2}}|s\rangle - \frac{5}{4\sqrt{2}}|101\rangle$ or using the $|\Psi\rangle = \frac{\sqrt{7}}{\sqrt{8}}|s\rangle + \frac{1}{\sqrt{8}}|101\rangle$ expression $|\Psi_3\rangle = \frac{1}{2}|\Psi\rangle - \frac{3}{2\sqrt{2}}|101\rangle$ because.

$$\frac{1}{2}|\Psi\rangle - \frac{3}{2\sqrt{2}}|101\rangle = \frac{\sqrt{7}}{4\sqrt{2}}|s\rangle + \left(\frac{1}{4\sqrt{2}} - \frac{3}{2\sqrt{2}}\right)|101\rangle = \frac{\sqrt{7}}{4\sqrt{2}}|s\rangle - \frac{5}{4\sqrt{2}}|101\rangle = |\Psi_3\rangle.$$

The last step is:

$$|\Psi_4\rangle = (2|\Psi\rangle\langle\Psi| - I)|\Psi_3\rangle = (2|\Psi\rangle\langle\Psi| - I)\left(\frac{1}{2}|\Psi\rangle - \frac{3}{2\sqrt{2}}|101\rangle\right) =$$

$$= 2 \cdot \frac{1}{2}|\Psi\rangle\langle\Psi\|\Psi\rangle - 2 \cdot \frac{3}{2\sqrt{2}}|\Psi\rangle\langle\Psi\|101\rangle - \frac{1}{2}|\Psi\rangle + \frac{3}{2\sqrt{2}}|101\rangle =$$

$$= |\Psi\rangle - \frac{3}{2\sqrt{2}}|\Psi\rangle \cdot \frac{1}{2\sqrt{2}} - \frac{1}{2}|\Psi\rangle + \frac{3}{2\sqrt{2}}|101\rangle = -\frac{1}{4}|\Psi\rangle + \frac{3}{2\sqrt{2}}|101\rangle$$

Using the $|\Psi\rangle = \frac{\sqrt{7}}{2\sqrt{2}}|s\rangle + \frac{1}{2\sqrt{2}}|101\rangle$ expression:

$$|\Psi_4\rangle = |\Psi\rangle = -\frac{\sqrt{7}}{8\sqrt{2}}|s\rangle - \frac{1}{8\sqrt{2}}|101\rangle + \frac{3}{2\sqrt{2}}|101\rangle = -\frac{\sqrt{7}}{8\sqrt{2}}|s\rangle + \frac{11}{8\sqrt{2}}|101\rangle.$$

A measurement of the state $|\Psi_4\rangle$ in the computational basis will project it into state $|101\rangle$ with probability: $p = \left|\frac{11}{8\sqrt{2}}\right|^2 \cong \left|\frac{121}{128}\right| \cong 0,9453$.

The basic idea of Grover iteration became obvious through this example: it increases the amplitude of the state that carry the desired information and decreases the amplitude of an other state $|i\rangle$.

In what follows we will show how important is to know the number of repetition of Grover iteration.

Let us suppose that, we don't know the value of *k*, and continue to perform further Grover iterations:

$$|\Psi_5\rangle = -\frac{\sqrt{7}}{8\sqrt{2}}|s\rangle - \frac{11}{8\sqrt{2}}|101\rangle \text{ or using } |\Psi\rangle = \frac{\sqrt{7}}{2\sqrt{2}}|s\rangle + \frac{1}{2\sqrt{2}}|101\rangle:$$

$$|\Psi_5\rangle = -\frac{1}{4}|\Psi\rangle - \frac{5}{4\sqrt{2}}|101\rangle.$$

$$|\Psi_6\rangle = (2|\Psi\rangle\langle\Psi| - I)|\Psi_5\rangle = (2|\Psi\rangle\langle\Psi| - I)\left(-\frac{1}{4}|\Psi\rangle - \frac{5}{4\sqrt{2}}|101\rangle\right) =$$

$$= -\frac{7}{8}|\Psi\rangle + \frac{5}{4\sqrt{2}}|101\rangle = -\frac{7}{8}\left(\frac{\sqrt{7}}{2\sqrt{2}}|s\rangle + \frac{1}{2\sqrt{2}}|101\rangle\right) + \frac{5}{4\sqrt{2}}|101\rangle \Rightarrow$$

$$\Rightarrow |\Psi_6\rangle = -\frac{7\sqrt{7}}{16\sqrt{2}}|s\rangle + \frac{13}{16\sqrt{2}}|101\rangle$$

If we measure now the state $|\Psi_6\rangle$, we et the result $|101\rangle$ with probability:

$$p = \left|\frac{13}{16\sqrt{2}}\right|^2 = \left|\frac{169}{512}\right| \cong 0{,}33, \text{ and some other state with probability:}$$

$$p' = \left|-\frac{7\sqrt{7}}{16\sqrt{2}}\right|^2 = \left|\frac{343}{512}\right| \cong 0{,}67.$$

So, if we continue the Grover iteration, after $k = \text{round}\left(\frac{\pi}{4}\sqrt{N}\right)$ steps the probability of finding the marked state begins to decline while the possibility of error increases better and better.

**Implementation**

In the following we can reed the Grover's opinion about the implementation possibility of his algorithm:

"This algorithm is likely to be simpler to implement as compared to other quantum mechanical algorithms for the following reasons:

   (i)     The only operations required are, first, the Walsh-Hadamard transform, and second, the conditional phase shift operation both of which are relatively easy as compared to operations required for other quantum mechanical algorithms

(ii) Quantum mechanical algorithms based on the Walsh-Hadamard transform are likely to be much simpler to implement than those based on the "large scale Fourier transform".

(iii) The conditional phase shift would be much easier to implement if the algorithm was used in the mode where the function at each point was computed rather than retrieved form memory. This would eliminate the storage requirements in quantum memory.

(iv) In case the elements had to be retrieved from a table (instead of being computed as discussed in (iii)), in principle it should be possible to store the data in classical memory and only the sampling system need be quantum mechanical. This is because only the system under consideration needs to undergo quantum mechanical interference, not the bits in the memory."

**Experimental realization using optical quantum computation technology**

We restrict our attention to the case where the number of elements in the search space is $N = 4$.

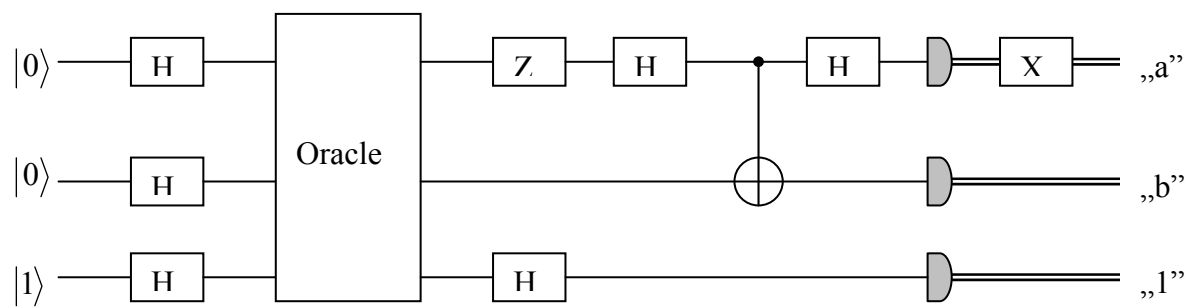

Fig. 1.

Fig. 1. is a circuit diagram for the four-element Grover algorithm. The top two qubits are the data qubits, initialised in state $||0\rangle|0\rangle$, while the bottom qubit is the oracle qubit, initialised in state $|1\rangle$. The boxes labelled H and Z represent the one-qubit Hadamard and Pauli Z gates, respectively. The CNOT is denoted by the usual symbol, while the grey half-circles represent one-qubit measurements in the computational basis, whose output appears on the classical output wires (double lines). The final X gate represents the classical NOT required to put the output into the correct form. The measurement always gives "1" on the oracle qubit, while the data qubits give "a" and "b". It is straightforward to show that, in principle, *ab* is the state marked by the oracle.

It can be verified directly that this circuit, using only one oracle call, gives the correct answer with probability 1, compared to the average of 2.25 oracle calls that must be made with a classical circuit. For example, if the solution is 11, then the output of the circuit is a = 1 and b = 1.

**Implementing the oracle**

An oracle is a quantum circuit that recognizes solutions to a given problem. For example, suppose we wish to solve a version of the travelling salesman problem, where the goal is to find a route visiting a given collection of cities that is shorter than some specified length L. Although it is in general hard to find such a route, it is easy to recognize whether a proposed route solves the problem: simply add up the total distance the salesman would travel on the proposed route, and compare it to L. Specifically, an oracle is a circuit that, given an input consisting of a potential solution to a problem, flips the sign of an ancilla qubit if and only if the input is a solution to the problem. Since the only action of the oracle is to recognize solutions, its internal structure is unimportant in a test of the algorithm itself. Thus, for our purposes, the choice of oracle is arbitrary, and may be chosen to be as simple as possible.

Although the internal workings of the oracle are unimportant for the purposes of testing the algorithm, the complexity of implementing some oracle must be included to characterize the difficulty of performing the experiment. A simple implementation of an oracle marking one of four states is a Toffoli gate, with the control qubits negated where necessary to specify any of the states 00, 01, 10, or 11 (see the Fig. 2a. for the example where the marked state is 11).

If the marked state is 11, the action of the oracle on the three qubits is to take the state $(|00\rangle + |01\rangle + |10\rangle + |11\rangle)(|0\rangle - |1\rangle)$ to $(|00\rangle + |01\rangle + |10\rangle)(|0\rangle - |1\rangle) + |11\rangle(|1\rangle - |0\rangle)$ (omitting the normalization). Thus the oracle simply has the effect of flipping the sign of the marked state. The ancilla is not used again, so it can be discarded at this point.

Toffoli gates are difficult to implement in LOQC because there is no known way to implement one without using several CNOT gate. However, for our purposes, a full Toffoli gate is not required because the ancilla qubit plays such a limited role.

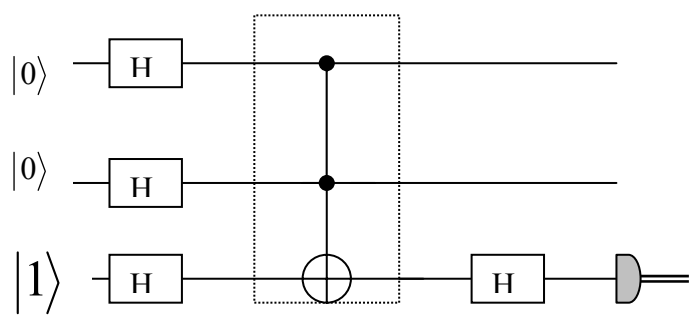

Fig. 2a.

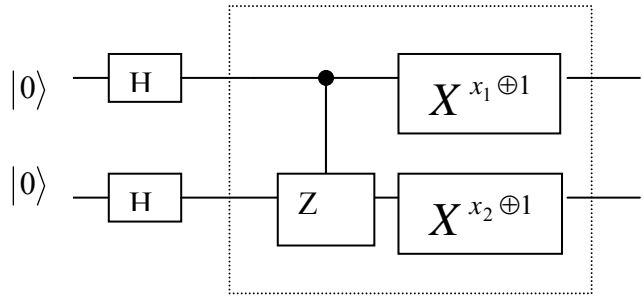

Fig. 2b.

FIG. 2a: The circuit shows the beginning of the Grover circuit with an example oracle (inside the dashed box) marking the item 11. We have implemented the oracle using a variant of the Toffoli gate, where the state of the third qubit is flipped when the first two qubits are in the state $|11\rangle$, as indicated by the closed circles on the control qubits. We have moved the measurement on the third qubit forward since it plays no further role in the algorithm. However, this circuit is in fact equivalent to the simplification in the Fig 2b., where the Toffoli has been replaced by a controlled-Z operation followed by the X gates

A simplified circuit to implement the four-element Grover algorithm is given in Fig. 3.

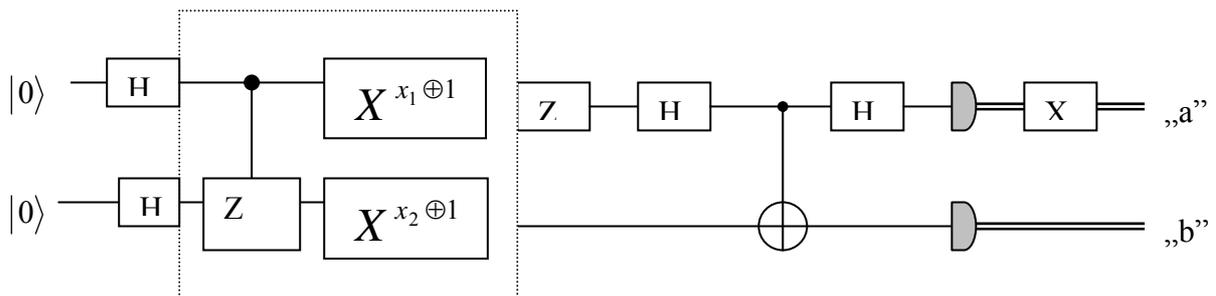

Fig. 3.

FIG. 3: Inserting the simplified oracle of Fig. 2b. into the circuit of Fig. 1 gives this circuit. Note that the marked state is specified inside the oracle (the dashed box) by the values of $x_1$ and $x_2$ used to determine whether or not the X gates are applied Under ideal circumstances, the output of the circuit is a = $x_1$ and b = $x_2$.

We are verifying directly that the circuit in fig 3. gives the correct answer to the search problem using only one oracle call.

In our example, if the solution is 11, then the output of the circuit is $a = 1$ and $b = 1$.

— Initial state is: $(H|0\rangle)^{\otimes 2} = \frac{1}{2}(|00\rangle + |01\rangle + |10\rangle + |11\rangle)$; which is the oracle input

— In this special case the oracle makes just one transformation- the Controlled –Z -,in order to invert the sign of the marked state because $x_1 = x_2 = 1$, and $X^{x_1 \oplus 1} = X^{x_2 \oplus 1} = I$, where *I* is the identity:

$$\frac{1}{2}(|00\rangle + |01\rangle + |10\rangle + |11\rangle) \xrightarrow{C-Z} \frac{1}{2}(|00\rangle + |01\rangle + |10\rangle - |11\rangle)$$

— The following two transformations: Z and H, are performed just on the first qubit ($x_1$):

$$\frac{1}{2}(|00\rangle + |01\rangle + |10\rangle - |11\rangle) \xrightarrow{Z \to x_1} \frac{1}{2}(|00\rangle + |01\rangle - |10\rangle + |11\rangle) \xrightarrow{H \to x_1}$$

$$\frac{1}{2\sqrt{2}}((|0\rangle + |1\rangle)(|0\rangle + |1\rangle) + (|0\rangle - |1\rangle)(|1\rangle - |0\rangle)) = \frac{1}{\sqrt{2}}(|01\rangle + |10\rangle)$$

— In the next step the C-NOT gate operates on both qubit as follows: it changes the second bit if the first bit is 1 and leaves this bit unchanged otherwise:

$$\frac{1}{\sqrt{2}}(|01\rangle + |10\rangle) \xrightarrow{C-NOT} \frac{1}{\sqrt{2}}(|01\rangle + |11\rangle)$$

— Repeat the H transformation on the first qubit:

$$\frac{1}{\sqrt{2}}(|01\rangle + |11\rangle) \xrightarrow{H \to x_1} \frac{1}{\sqrt{2}}\left(\left(\frac{1}{\sqrt{2}}|0\rangle + |1\rangle\right)|1\rangle + \left(\frac{1}{\sqrt{2}}|0\rangle - |1\rangle\right)|1\rangle\right) =$$

$$= \frac{1}{2}(|01\rangle + |11\rangle + |01\rangle - |11\rangle) = |01\rangle$$

— We measure the qubits in the computational basis separately, and perform an X (classical NOT) transformation on the first bit:

$$|01\rangle \xrightarrow{Measurement} \begin{cases} 0 \xrightarrow{X} 1 = "a" \\ 1 \to 1 = "b" \end{cases}, \text{ the output is: } ab=11$$

**Notations**

— **H** is the Hadamard gate defined as: $H = \frac{1}{\sqrt{2}}\begin{bmatrix} 1 & 1 \\ 1 & -1 \end{bmatrix}$,

more precisely:
$$H: \begin{aligned} |0\rangle &\to \frac{1}{\sqrt{2}}(|0\rangle + |1\rangle) \\ |1\rangle &\to \frac{1}{\sqrt{2}}(|0\rangle - |1\rangle) \end{aligned}$$

— **X** is the classical NOT gate, or Pauli X transformation defined as:

$$X = \begin{pmatrix} 0 & 1 \\ 1 & 0 \end{pmatrix}, \text{ more precisely: } \quad X: \begin{aligned} |0\rangle &\to |0\rangle \\ |1\rangle &\to |1\rangle \end{aligned}$$

— **Z** is the phase shift gate, or Pauli Z transformation defined as:

$$Z = \begin{pmatrix} 1 & 0 \\ 0 & -1 \end{pmatrix} \text{ or more precisely: } \quad Z: \begin{aligned} |0\rangle &\to |0\rangle \\ |1\rangle &\to -|1\rangle \end{aligned}$$

— **CNOT** is the controlled NOT gate, which applies a NOT on the second bit, called the target bit if the first control bit is 1:

$$CNOT = \begin{pmatrix} 1 & 0 & 0 & 0 \\ 0 & 1 & 0 & 0 \\ 0 & 0 & 0 & 1 \\ 0 & 0 & 1 & 0 \end{pmatrix} \quad \text{or:} \quad \begin{array}{l} CNOT: |00\rangle \to |00\rangle \\ |01\rangle \to |01\rangle \\ |10\rangle \to |11\rangle \\ |11\rangle \to |10\rangle \end{array}$$

— **C-Z** is the controlled **Z** gate, which applies the Z transformation on the second bit, t if the first control bit is 1:

Matrix representation: $C-Z = \begin{pmatrix} 1 & 0 & 0 & 0 \\ 0 & 1 & 0 & 0 \\ 0 & 0 & 1 & 0 \\ 0 & 0 & 0 & -1 \end{pmatrix}$, or: $\begin{array}{l} C-Z: |00\rangle \to |00\rangle \\ |01\rangle \to |01\rangle \\ |10\rangle \to |10\rangle \\ |11\rangle \to -|11\rangle \end{array}$

— Toffoli gate or controlled–controlled–NOT, which negates the last bit of three if and only if the first two are both 1. matrix representation: $\begin{pmatrix} I_{4x4} & \vdots & 0 \\ \cdots & \cdots & \cdots \\ 0 & \vdots & CNOT \end{pmatrix}$

**References**

1. M.A. Nielsen and I.L. Chuang, Quantum Computation and Quantum Information, Cambridge University Press, Cambridge (2000).
2. J. Preskill, Quantum Information and Computation, Lecture Notes, California Institute of Technology (1998).
3. G. Brassard, P. Høyer, and A. Tapp, Quantum Counting, quant-ph/9805082.
4. .Grover, Lov K., A framework for fast quantum mechanical algorithms, quant-ph/9711043.
5. Grover, L., Quantum mechanics helps in searching for a needle in a haystack (quant-ph/9605043)
6. Grover, L., Phys. Rev. Lett. 78, (1997), pp 325 - 328.
7. Gruska, Jozef, "Quantum Computing," McGraw-Hill, (1999)
8. Jozsa, Richard, Searching in Grover's Algorithm, quant-ph/9901021.
9. Andris Ambainis. Quantum query algorithms and lower bounds
10. P. G. Kwiat, J. R. Mitchell, P. D. D. Schwindt, and A. G. White, Grover's search algorithm: An optical approach, quant- ph/ 9905086
11. Christof Zalka, Could Grover's quantum algorithm help in

searching an actual database? quant- ph/ 9901068

12. Ofer Biham, Daniel Shapira and Yishai Shimoni, Analysis of Grover's quantum search algorithm as a dynamical system, quant- ph/ 0307141

13. Jennifer L. Dodd, Timothy C. Ralph and G. J. Milburn, Grover's algorithm in optical quantum computation

14. J´er´emie Roland and Nicolas J. Cerf, Quantum circuit implementation of the Hamiltonian versions of Grover's algorithm, quant- ph/ 0302138

15. Manoj K. Samal and Partha Ghose, Grover's search algorithm and the quantum measurement problem, quant- ph/ 0202176

16. Daniel Braun, Quantum Chaos and Quantum Algorithms, quant- ph/ 0110037

17. Goong Chen and Shunhua Sun , Generalization of Grover's Algorithm to Multiobject Search in Quantum Computing, General Unitary Transformati